\documentclass[12pt]{article}
\usepackage[utf8]{inputenc}
\usepackage{amsmath}
\usepackage{amsfonts}
\usepackage{amssymb}
\usepackage{extsizes}
\usepackage{graphicx}
\usepackage{tikz}
\usepackage[margin=0.75in]{geometry}
\usepackage{physics}
\def\vecsign{\mathchar"017E}
\def\dvecsign{\smash{\stackon[-1.95pt]{\vecsign}{\rotatebox{180}{$\vecsign$}}}}
\def\dvec#1{\def\useanchorwidth{T}\stackon[-4.2pt]{#1}{\,\dvecsign}}
\usepackage{stackengine}
\stackMath

\begin{document}
	
	\title{\textbf{Noether type formulation for space dependent polynomial symmetries}}
	\author{Rabin Banerjee\\
	\emph{S. N. Bose National Centre for Basic Sciences}, \\
	\emph{JD Block, Sector III, Salt Lake, Kolkata 700106, India}}
\date{}	
\maketitle	

	\begin{abstract}

	We develop a systematic algorithm, based on Noether's theorem, for defining the various currents in theories invariant under space dependent polynomial symmetries. A master equation is given that yields all the conservation laws corresponding to these currents. Explicit demonstration has been provided for dipole and quadrupole conservation symmetries.

		\end{abstract}


		\section{Introduction}		For nearly a century, the search for novel quantum states of matter has been an important and interesting scientific pursuit, both theoretically \cite{chamon,Wen,A,kitaev} and experimentally \cite{nat,sci,exp3}. Theoretical interests in novel classes of quantum states of matter that do not fit well into any current framework have recently grown \cite{chamon,Wen,Vijay,sondhi}. Among these are the fracton \cite{Vijay,Nandkishore0} phases, and they represent a new frontier of quantum  physics.  In contrast to other (quasi) particles, quasi-excitation fractons are defined by their extremely low mobility \cite{Vijay,Nandkishore2}. As a result of their immobility, fractons are extraordinarily slow to respond to an applied electric field. Fracton composites, on the other hand, can move freely. On the theoretical side, the fracton's immobility in an isolated state and mobility in the presence of additional fractons is a strange phenomenon \cite{ Nandkishore0,Nandkishore2, gromov}. Clearly, this fracton phase of matter reveals an unknown (perhaps enlarged, beyond usual space-time) symmetry  which one can pursue to investigate using quantum field theoretic methods \cite{gromov, pretko1,Seiberg1,Seiberg2,bulmash, RB1,wang,yuan}.  An essential tool is to exploit, apart from charge conservation, higher order conservation laws related to dipole, quadrupole and other n-pole symmetries \cite{hoyos}. These are usually obtained by certain assumptions \cite{Seiberg1} or by inspection \cite{gromov, hoyos}.
		
		
		In the present letter we give a systematic algorithm, based on the Noether approach, to derive all the multiple currents and  conservation laws related to an n-pole symmetry. The parameter related to this symmetry is no longer global. Rather, it is a polynomial of ($n-1$)th order in  spatial coordinates generating the various shift symmetries \cite{horava}. Thus we are able to generalise the usual Noether approach, based on the first theorem, that is only valid for a global symmetry leading to charge conservation.\\

		We briefly introduce the Schroedinger theory, that has charge conservation, in a way that is amenable to the generalisation that we are looking for. This is followed by analysing, in successive sections, theories with linear, quadratic and higher shift symmetries. Concrete realisations are provided.

\section{Schroedinger Theory}
The nonrelativistic Schroedinger theory is first discussed that illuminates the various facets to be subsequently generalised.
Consider an arbitrary Lagrangian $\mathcal{L}(\Phi, \partial_{0}\Phi,\partial_{i}\Phi)$, where $\Phi$ is a complex scalar, that has a global $U(1)$ symmetry,
\begin{equation}\label{1}
\Phi\rightarrow e^{i\alpha}\Phi~~,~~\Phi^*\rightarrow e^{-i\alpha} \Phi^*
\end{equation}
Taking the infinitesimal version, $\delta\Phi=i\alpha\Phi$, the invariance of the Lagrangian requires,
\begin{eqnarray}\label{quadvarL}
	\delta\mathcal{L}=i\alpha\frac{\partial\mathcal{L}}{\partial\Phi}\Phi+i\alpha\frac{\partial\mathcal{L}}{\partial(\partial_{0}\Phi)}\partial_0 \Phi+i\alpha\frac{\partial\mathcal{L}}{\partial(\partial_{i}\Phi)}\partial_i \Phi +c.c.=0
\end{eqnarray}
where $c.c.$ denotes complex conjugation.\\
The simplest (quadratic) Lagrangian satisfying (\ref{quadvarL}) is just the Schroedinger theory,
\begin{eqnarray}\label{quadL}
	\mathcal{L}=\frac{i}{2}\Phi^* \dvec{\partial}_0\Phi-\frac{1}{2m}\partial_i\Phi^*\partial_i\Phi 
\end{eqnarray}
with the global symmetry leading to the conservation law,
\begin{eqnarray}\label{1stconserv}
\partial_0j_0-\partial_ij_i=0
\end{eqnarray}
where the currents are defined by Noether's first theorem,
\begin{eqnarray}\label{Noether1}
\alpha j^0=\frac{\partial\mathcal{L}}{\partial(\partial_{0}\Phi)}\delta\Phi+c.c.~~,~~\alpha j^i=\frac{\partial\mathcal{L}}{\partial(\partial_{i}\Phi)}\delta\Phi+c.c.
\end{eqnarray}
By using their  explicit structures,
\begin{eqnarray}\label{explicitcurr}
j^0=-j_0=-\Phi^*\Phi~~;~~j^i=j_i=\frac{i}{2m}(\Phi^*\partial_i\Phi-\Phi\partial_i\Phi^*)
\end{eqnarray}
the conservation law (\ref{1stconserv}) is easily verified, on shell. Conservation of charge  $Q=\int_{space} j_0$ also follows from (\ref{1stconserv}).

The above discussion crucially hinges on the global nature of the parameter $\alpha$. If $\alpha$ is taken to be local (i.e. spacetime dependent) then the Lagrangian (\ref{quadL}) is no longer invariant and the analysis breaks down. In the next section we show how to modify the Noether procedure  for a space dependent parameter. 

\section{Theory with space-dependent linear shift symmetry}
As a non-trivial example we try to pursue a similar analysis where the Lagrangian is invariant under (\ref{1}), but now with a space-dependent linear shift,
\begin{equation}\label{defalpha}
\alpha=\alpha_0+\alpha_ix_i
\end{equation}
with constant $\alpha_0$ and $\alpha_i$.
Invariance under the above shift symmetry for a Lagrangian that is a function of its fields and first derivatives requires the following conditions,
\begin{subequations}\label{38}
\begin{eqnarray}\label{cond1a}
\frac{\partial\mathcal{L}}{\partial\Phi}(i\Phi)+\frac{\partial\mathcal{L}}{\partial(\partial_{0}\Phi)}(i\partial_0\Phi)+\frac{\partial\mathcal{L}}{\partial(\partial_{i}\Phi)}(i\partial_i\Phi)+c.c.=0
\end{eqnarray}
\begin{equation}
\frac{\partial\mathcal{L}}{\partial(\partial_{i}\Phi)}(i\Phi)+c.c.=0
\end{equation}
\end{subequations}
Both conditions cannot be satisfied simultaneously.\footnote{If we relax the conditions modulo a total derivative so that the action is kept invariant, then these can be satisfied. We discuss this later when considering  explicit realisations in section 3.1} Hence a more general Lagrangian has to be taken. It turns out that it must have at least two space derivatives. Then, invariance under (\ref{1}) demands\footnote{The temporal contribution is not explicitly written since the shift symmetry involves only spatial terms.},
\begin{subequations}\label{39}
\begin{equation}\label{demand1}
i(\alpha_0+\alpha_kx_k)\bigg[\frac{\partial\mathcal{L}}{\partial\Phi}\Phi+\frac{\partial\mathcal{L}}{\partial(\partial_{i}\Phi)}(\partial_i\Phi)+\frac{\partial\mathcal{L}}{\partial(\partial_i\partial_{j}\Phi)}(\partial_i\partial_j\Phi)\bigg]+c.c.=0
\end{equation}
\begin{equation}\label{demand1b}
 \partial_i\alpha\bigg[\frac{\partial\mathcal{L}}{\partial(\partial_i\Phi)}\Phi+2\frac{\partial\mathcal{L}}{\partial(\partial_i\partial_{j}\Phi)}(\partial_j\Phi)\bigg]+c.c.=0   
\end{equation}
\end{subequations}
While the first is associated with usual charge invariance, the second is new and connected with dipole symmetry, which will be transparent as we go on.

Noether's first theorem for a Lagrangian involving second derivatives leads to a conserved current,
\begin{equation}\label{one-currdefn}
\alpha j_i=\frac{\partial\mathcal{L}}{\partial(\partial_i\Phi)}\delta\Phi+\frac{\partial\mathcal{L}}{\partial(\partial_i\partial_{j}\Phi)}(\partial_j\delta\Phi)-\bigg(\partial_j\frac{\partial\mathcal{L}}{\partial(\partial_i\partial_{j}\Phi)}\bigg)\delta\Phi+c.c.
\end{equation}
where $\alpha$ is a constant. For the linearly shifted parameter $\alpha$ (\ref{defalpha}), we posit the definition,
\begin{equation}\label{two-currdefn}
\alpha j_i-(\partial_j\alpha) j_{ji}=\frac{\partial\mathcal{L}}{\partial(\partial_i\Phi)}\delta\Phi+\frac{\partial\mathcal{L}}{\partial(\partial_i\partial_{j}\Phi)}(\partial_j\delta\Phi)-\bigg(\partial_j\frac{\partial\mathcal{L}}{\partial(\partial_i\partial_{j}\Phi)}\bigg)\delta\Phi+c.c.
\end{equation}
This generalisation of the Noether current for a linear shift symmetry is dictated by the fact that, for a constant (global) $\alpha$, it should reduce to the usual definition; thus the correction must involve a linear derivative. By appropriate contraction, a tensor current is introduced to achieve consistency. We show it yields a viable definition.  After an appropriate reshuffling of derivatives, (\ref{two-currdefn}) is expressed as,
\begin{equation}\label{two-current}
\alpha j_i-(\partial_j\alpha) j_{ji}=\frac{\partial\mathcal{L}}{\partial(\partial_i\Phi)}\delta\Phi+2\frac{\partial\mathcal{L}}{\partial(\partial_i\partial_{j}\Phi)}(\partial_j\delta\Phi)-\partial_j\bigg(\frac{\partial\mathcal{L}}{\partial(\partial_i\partial_{j}\Phi)}\delta\Phi\bigg)+c.c.
\end{equation}
Using $\delta\Phi$ from (\ref{1}), the condition (\ref{demand1b}),
and equating either side of (\ref{two-current}) yields,
\begin{equation}\label{jiandjij}
j_i=\partial_j j_{ji}
\end{equation}
where,
\begin{equation}\label{2ndrankc}
j_{ji}=-i\bigg[\frac{\partial\mathcal{L}}{\partial(\partial_i\partial_{j}\Phi)}\Phi\bigg]
\end{equation}
The temporal part remains unaffected,
\begin{equation}\label{temporal}
\alpha j_0=(\alpha_0+\alpha_kx_k)j_0=\frac{\partial\mathcal{L}}{\partial(\partial_0\Phi)}\delta\Phi
\end{equation}
Relations (\ref{two-current}) and (\ref{temporal}) are used to deduce conservation laws analogous to (\ref{1stconserv}),
\begin{equation}\label{2ndconserv}
\partial_0(\alpha j_0)-\partial_i[\alpha j_i-(\partial_j \alpha) j_{ji}]=0
\end{equation}
which yields,
\begin{equation}\label{2ndcond}
\alpha[\partial_0j_0-\partial_ij_i]-\partial_i \alpha j_i+\partial_j \alpha \partial_ij_{ji}=0
\end{equation}
Using (\ref{jiandjij}) leads to the conservation law,
\begin{eqnarray}\label{currentconserve}
\partial_0j_0-\partial_ij_i=0
\end{eqnarray}
This derivation of the charge conservation hides the fact that there are two conservation laws, related to charge and dipole symmetry associated with the parameters $\alpha_0$ and $\alpha_k$ respectively. We thus equate coefficients of $\alpha_0$ and $\alpha_k$ separately from (\ref{2ndconserv}). From $\alpha_0$, the standard current conservation (\ref{currentconserve}) is again reproduced. Equating coefficients of $\alpha_k$ gives,
\begin{eqnarray}\label{dipole1}
\partial_0J_0^k-\partial_iJ_i^k=0
\end{eqnarray}
where the new currents are,
\begin{subequations}\label{newcurrents}
\begin{eqnarray}\label{currentcapital}
J_0^k=x_kj_0
\end{eqnarray}
\begin{eqnarray}
J_i^k=x_kj_i-j_{ki}=x_k\partial_jj_{ji}-j_{ki}
\end{eqnarray}
\end{subequations}

All relations (\ref{jiandjij}), (\ref{2ndrankc}), (\ref{currentconserve}), (\ref{dipole1}) and (\ref{newcurrents}) follow from the definition (\ref{two-currdefn}) without any additional inputs. This shows the consistency of the formulation.

Relation (\ref{dipole1}) yields the familiar dipole conservation law,
\begin{eqnarray}\label{Dipolelaw}
Q^k=\int_{space}x^kj_0=\int_{space}J_0^k~~~; ~~\dot{Q}^k=0
\end{eqnarray}
Since tensor sources $j_{ij}$ have appeared, it implies that these theories admit symmetric tensor gauge fields. Their gauge transformation properties are determined by using  relation (\ref{jiandjij}) to connect usual vector couplings with tensor couplings,
\begin{equation}\label{duality1}
j_iA_i= \partial_jj_{ji}A_{i}= j_{ij}A_{ij}
\end{equation}
where a reshuffling of derivatives yields, 
\begin{equation}
A_{ij}=-\frac{1}{2}(\partial_iA_j+\partial_jA_i)
\end{equation}
 Usual gauge transformations for $A_{i}$,
\begin{equation}
A_i\rightarrow A_{i}+\partial_i\beta
\end{equation}
gives the gauge transformation for $A_{ij}$,
\begin{equation}
A_{ij}\rightarrow A_{ij}-\partial_i\partial_j\beta
\end{equation}
where $\beta$ is the gauge parameter. The matter sector can be coupled to usual gauge fields with the standard Maxwell Lagrangian, or equivalently, to symmetric tensor gauge fields with a nonstandard Lagrangian \cite{pretko1, RB1}.

\subsection{Explicit realisations}

Although an explicit realisation of a theory with charge and dipole conservation has been given in \cite{pretko1, wang, yuan} using fracton gauge principle or an appropriate change of variables \cite{Seiberg1}, it is nice to reproduce this result in our approach and see its consequences.

It is  possible to construct a viable Lagrangian, depending on two spatial derivatives so that it satisfies the conditions (\ref{39}). Especially (\ref{demand1b}) immediately suggests the following structure,
\begin{equation}\label{L2nd-order}
    \mathcal{L}_t=t|\Phi\partial_i\partial_j\Phi-\partial_i\Phi\partial_j\Phi|^2
    \end{equation}
    where $t$ is just a normalising factor. This Lagrangian trivially satisfies (\ref{demand1}) also. By using (\ref{2ndrankc}) we obtain,
    \begin{eqnarray}
j_{ij}=-i\bigg(\Phi^2(\Phi^*\partial_i\partial_j\Phi^*-\partial_i\Phi^*\partial_j\Phi^*)-{\Phi^*}^2(\Phi\partial_i\partial_j\Phi-\partial_i\Phi\partial_j\Phi)\bigg)
    \end{eqnarray}
   The above current involves derivatives and transforms nontrivially under rotations. \\
   
   A rotationally invariant current, one that does not involve derivatives, can also be found. In this case  consider a Lagrangian that involves only single derivatives,
  \begin{equation}
       \mathcal{L}_u=u\, i\Big({\Phi^*}^2 \partial_i\Phi \partial_i\Phi - \Phi^2\partial_i\Phi^* \partial_i\Phi^*\Big)
  \end{equation}
   This Lagrangian is akin to conventional theories, not involving higher derivatives. To see if it has both charge and dipole symmetries, we have to now look for the conditions (\ref{38}) since these are the relevant ones for this case. We find that the first one, expectedly, is satisfied trivially,  while the second leads to,
   \begin{equation}
\frac{\partial\mathcal{L}}{\partial(\partial_{i}\Phi)}(i\Phi)+c.c.= -\partial_i(|\Phi|^4)
\end{equation}
Thus, if we allow for quasi invariance (the action being invariant), then this is also an allowed form. 

The Noether formulation has  to account for this boundary term, apart from the fact that only linear derivatives appear.  In this case, therefore, the Noether definition (\ref{two-currdefn}) is modified as,
\begin{equation}\label{two-currdefn1}
\alpha j_i-(\partial_j\alpha) j_{ji}=\frac{\partial\mathcal{L}}{\partial(\partial_i\Phi)}\delta\Phi+c.c.+ \alpha_i |\Phi|^4
\end{equation}
Comparing coefficients of $\alpha$ and $\alpha_i$, one is immediately led to the results,
\begin{equation}
j_{ij}=-\delta_{ij}|\Phi|^4\,\,;\,\,j_i=-\partial_i |\Phi|^4=\partial_j j_{ji}
\end{equation}
 The tensor current is rotationally invariant and is connected to the vector current in the same manner as discussed earlier for theories satisfying both charge and dipole symmetries.

    We next discuss the example of shift symmetry involving quadratic terms.
    
    \section{Theory with space-dependent quadratic shift symmetry}
    The analysis in the previous section is now extended to include quadratic terms in the shift symmetry,
    \begin{equation}\label{shift2nd}
    \alpha=\alpha_0+\alpha_ix_i+\frac{1}{2}\alpha_{ij}x_ix_j
\end{equation}

Following previous arguments it is seen that the Lagrangian must contain terms with triple derivatives. Then invariance of the Lagrangian $\mathcal{L}(\Phi, \partial_0\Phi, \partial_{i}\Phi, \partial_{i}\partial_{j}\Phi,\partial_{i}\partial_{j}\partial_k\Phi)$ under the infinitesimal transformation $\delta\Phi=i\alpha\Phi$ requires,
\begin{subequations}\label{cond3}
\begin{equation}\label{three-currenta}
i\bigg[\frac{\partial\mathcal{L}}{\partial\Phi}\Phi+\frac{\partial\mathcal{L}}{\partial(\partial_i\Phi)}\partial_i\Phi+\frac{\partial\mathcal{L}}{\partial(\partial_i\partial_{j}\Phi)}\partial_i\partial_j\Phi+\frac{\partial\mathcal{L}}{\partial(\partial_i\partial_{j}\partial_k\Phi)}\partial_i\partial_j\partial_k\Phi\bigg]+c.c.=0
\end{equation}
\begin{equation}\label{three-currentb}
i\bigg[\frac{\partial\mathcal{L}}{\partial(\partial_i\Phi)}\Phi+2\frac{\partial\mathcal{L}}{\partial(\partial_i\partial_{j}\Phi)}(\partial_j\Phi)+3\frac{\partial\mathcal{L}}{\partial(\partial_i\partial_{j}\partial_k\Phi)}\partial_j\partial_k\Phi\bigg]+c.c.=0
\end{equation}
\begin{equation}\label{three-currentc}
i\bigg[\frac{\partial\mathcal{L}}{\partial(\partial_i\partial_j\Phi)}\Phi+3\frac{\partial\mathcal{L}}{\partial(\partial_i\partial_{j}\partial_k\Phi)}\partial_k\Phi\bigg]+c.c.=0
\end{equation}
\end{subequations}
These conditions are essential for deriving conservation laws related to charge, dipole and quadrupole symmetries, respectively.
The generalised Noether current for this system with triple derivatives is given by following the same logic used in writing (\ref{two-currdefn}),
\small
\begin{eqnarray}\label{Noether3rd}
\alpha j_i-(\partial_j\alpha) j_{ji} +(\partial_j\partial_k\alpha)j_{jki}&=&\frac{\partial\mathcal{L}}{\partial(\partial_{i}\Phi)}\delta \Phi-\partial_{j}\bigg(\frac{\partial\mathcal{L}}{\partial(\partial_{i}\partial_{j}\Phi)}\bigg)\delta \Phi+\frac{\partial\mathcal{L}}{\partial(\partial_{i}\partial_{j}\Phi)}\partial_{j}\delta \Phi+\partial_{j}\partial_{k}\bigg(\frac{\partial\mathcal{L}}{\partial(\partial_{i}\partial_{j}\partial_{k}\Phi)}\bigg)\delta\Phi \nonumber \\
&+&\frac{{\partial\mathcal{L}}}{\partial(\partial_{i}\partial_{j}\partial_{k}\Phi)}\partial_{j}\partial_{k}\delta \Phi -\partial_{k}\bigg(\frac{\partial\mathcal{L}}{\partial(\partial_{i}\partial_{j}\partial_{k}\Phi)}\bigg) \partial_{j}\delta \Phi+ c.c. 
\end{eqnarray}
\normalsize
where, expectedly, double derivatives appear while rank 2 and rank 3 tensor currents are naturally introduced. After a rearrangement of the derivatives on the right hand side, the above equation is put in the following form:
\small
\begin{eqnarray}\label{condition3}
\alpha j_i-(\partial_j\alpha)j_{ji}+(\partial_j\partial_k\alpha)j_{kji}&=&\frac{\partial\mathcal{L}}{\partial(\partial_{i}\Phi)}\delta \Phi+2\frac{\partial\mathcal{L}}{\partial(\partial_{i}\partial_{j}\Phi)}\partial_{j}\delta \Phi+3\frac{{\partial\mathcal{L}}}{\partial(\partial_{i}\partial_{j}\partial_{k}\Phi)}\partial_{j}\partial_{k}\delta \Phi \nonumber \\
&-&\partial_{j}\bigg(\frac{\partial\mathcal{L}}{\partial(\partial_{i}\partial_{j}\Phi)}\delta \Phi +2\frac{{\partial\mathcal{L}}}{\partial(\partial_{i}\partial_{j}\partial_{k}\Phi)}\partial_{k}\delta \Phi\bigg)+\partial_{j}\partial_{k}\bigg(\frac{\partial\mathcal{L}}{\partial(\partial_{i}\partial_{j}\partial_{k}\Phi)}\delta\Phi\bigg)+c.c. \nonumber \\
\end{eqnarray}
\normalsize

The advantage of this form is that it is readily generalisable to higher orders, as will be discussed later. Further, the algebra simplifies   on using the conditions (\ref{three-currentb},\ref{three-currentc}). Carrying out the complete computation yields,
\begin{equation}\label{threecurrentdefn}
j_i=\partial_j\partial_kj_{jki} \,\,;\,\,  j_{ji}= \partial_kj_{kji}
\end{equation}
where,
\begin{equation}
j_{ijk}=i\frac{{\partial\mathcal{L}}}{\partial(\partial_{i}\partial_{j}\partial_{k}\Phi)} \Phi
\end{equation}
The tower of conservation laws may now be derived from (\ref{condition3}), after including the temporal contribution which remains unaffected,
\begin{eqnarray}\label{temporaleqn}
\partial_0(\alpha j_0)-\partial_i\bigg(\alpha j_i-(\partial_j\alpha)\partial_kj_{ijk}+(\partial_j\partial_k\alpha)j_{ijk}\bigg)=0
\end{eqnarray}
Factoring out $\alpha$ yields,
\begin{eqnarray}
\alpha(\partial_0j_0-\partial_ij_i)-\bigg((\partial_i\alpha) j_i-(\partial_i\partial_j\alpha)\partial_kj_{ijk}-(\partial_j\alpha)\partial_i\partial_kj_{ijk}+(\partial_j\partial_k\alpha)\partial_ij_{ijk}+(\partial_i\partial_j\partial_k\alpha)j_{ijk}\bigg)=0 \nonumber \\
\end{eqnarray}
Using (\ref{threecurrentdefn}) and the fact that $\alpha$ is a quadratic function in $x_i$ (\ref{shift2nd}), we conclude -
\begin{eqnarray}\label{chargeconserve3}
\partial_0j_0-\partial_ij_i=0
\end{eqnarray}
While this is an economical way of getting the basic (charge) conservation, it conceals the effect of the other two, namely  dipole and  quadrupole symmetries. These are obtained by independently considering the equalities related to $\alpha$, opening it by using (\ref{shift2nd}) and inserting in (\ref{temporaleqn}). Then $\alpha_0$ just leads to the charge conservation. Next comparing factors multiplying $\alpha_i$ gives,
\begin{eqnarray}\label{conserve3}
\partial_0(\alpha_jx_jj_0)-\partial_i(\alpha_jx_jj_i-\alpha_j\partial_kj_{ijk})=0
\end{eqnarray}
Defining new currents,
\begin{eqnarray}\label{newcurrents3}
J_{0j}=x_jj_0~~,~~J_{ji}=x_jj_i-\partial_kj_{ijk}=x_j\partial_l\partial_mj_{lmi}-\partial_kj_{ijk}
\end{eqnarray}
leads to the dipole conservation,
\begin{eqnarray}
\partial_0J_{0j}-\partial_iJ_{ji}=0
\end{eqnarray}
Finally, comparing factors multiplying $\alpha_{ij}$,
\begin{eqnarray}
\partial_0\bigg(\frac{1}{2}\alpha_{ij}x_ix_jj_0\bigg)-\partial_i\bigg(\frac{1}{2}\alpha_{jk}x_jx_kj_i-\frac{1}{2}\alpha_{jl}(x_l\partial_kj_{ijk}+x_j\partial_kj_{ilk})+\alpha_{jk}j_{jki}\bigg)=0
\end{eqnarray}
where the second term in the second parenthesis has been suitably symmetrised. Thus, the above equation provides the quadrupole conservation,
\begin{eqnarray}\label{quadcons}
\partial_0J_{ij0}-\partial_kJ_{kji}=0
\end{eqnarray}
where the new currents are defined by,
\begin{eqnarray}\label{timecurrent}
J_{ij0}=\frac{1}{2}x_ix_jj_0
\end{eqnarray}
and,
\begin{eqnarray}\label{spacecurrent}
J_{kji}=\frac{1}{2}x_ix_jj_k-\frac{1}{2}(x_i\partial_pj_{kjp}+x_j\partial_pj_{kip})+j_{jik}
=\frac{1}{2}\Big(x_ix_j\partial_p\partial_qj_{pqk}-x_i\partial_pj_{kjp} - x_j\partial_pj_{kip}\Big)+j_{jik}
\end{eqnarray}
It is easy to verify, by substituting (\ref{timecurrent}) and (\ref{spacecurrent}) in (\ref{quadcons}), that,
\begin{eqnarray}
x_ix_j\bigg(\partial_0j_0-\partial_kj_k\bigg)=0
\end{eqnarray}
on account of (\ref{chargeconserve3}) and serves as a consistency check.
\subsection{Explicit realisation}
We now provide an explicit realisation of a theory having charge, dipole and quadrupole conservations. A general term in the Lagrangian which will respect the three conservation laws will involve three spatial derivatives. Hence it can be written as,
\begin{eqnarray}\label{quadlag}
{{\cal L}_v}=	v|\Phi^2\partial_{i}\partial_{j}\partial_{k}\Phi+b\partial_{i}\Phi\partial_{j}\Phi\partial_{k}\Phi+c\Phi(\partial_{i}\Phi\partial_{j}\partial_{k}\Phi+\partial_{j}\Phi\partial_{k}\partial_{i}\Phi+\partial_{k}\Phi\partial_{i}\partial_{j}\Phi)|^2
\end{eqnarray}
where $b$ and  $c$ are constant real numbers. Note that the coefficient of the first term has been scaled to unity. These coefficients will be determined by using the conditions (\ref{cond3}). The first of these is trivially satisfied since it conforms to the basic charge conservation. Nontrivial consequences follow from the other two conditions. The last one, for instance, yields,
	\begin{eqnarray}
	& i(c+1)\bigg(|\Phi|^4\partial_{k}\Phi\partial_{ijk}\Phi^*+ b\Phi^2\partial_{k}\Phi\partial_{i}\Phi^*\partial_{j}\Phi^*\partial_{k}\Phi^* \nonumber \\
	&+|\Phi|^2\Phi\partial_{k}\Phi(\partial_{i}\Phi^*\partial_{j}\partial_{k}\Phi^*+\partial_{j}\Phi^*\partial_{k}\partial_{i}\Phi^*+\partial_{k}\Phi^*\partial_{i}\partial_{j}\Phi^*)\bigg)+c.c.=0
	\end{eqnarray}
which gives $c=-1$. Taking the other condition now gives $b=2$. This completely fixes the structure of the lagrangian {\footnote{After completing our paper we found  that a similar structure was given in \cite{kristan}, but without any explanation regarding the method of its obtention.}. The operator inside the modulus in (\ref{quadlag}) transforms covariantly under the polynomial shift symmetry \cite{kristan}. 
Finally, including other possible terms, the Lagrangian looks like,
\begin{eqnarray}\label{Lagrangian3rdorder}
	\mathcal{L}=& u|\Phi^{* 3}\bigg(\Phi^2\partial_{i}\partial_{j}\partial_{k}\Phi+2\partial_{i}\Phi\partial_{j}\Phi\partial_{k}\Phi-\Phi(\partial_{i}\Phi\partial_{j}\partial_{k}\Phi+\partial_{j}\Phi\partial_{k}\partial_{i}\Phi+\partial_{k}\Phi\partial_{i}\partial_{j}\Phi)\bigg)|^2 \nonumber \\
	&+ v|\Phi^2\partial_{i}\partial_{j}\partial_{k}\Phi+2\partial_{i}\Phi\partial_{j}\Phi\partial_{k}\Phi-\Phi(\partial_{i}\Phi\partial_{j}\partial_{k}\Phi+\partial_{j}\Phi\partial_{k}\partial_{i}\Phi+\partial_{k}\Phi\partial_{i}\partial_{j}\Phi)|^2 \nonumber \\
	& + t |\Phi^2\partial_{i}\partial_{j}\partial_{j}\Phi+2\partial_{i}\Phi\partial_{j}\Phi\partial_{j}\Phi-\Phi(\partial_{i}\Phi\partial_{j}\partial_{j}\Phi+\partial_{j}\Phi\partial_{j}\partial_{i}\Phi+\partial_{j}\Phi\partial_{i}\partial_{j}\Phi)|^2 -m^2\Phi^{\dagger}\Phi \nonumber \\
\end{eqnarray}
 Now, the various currents corresponding to the \textbf{$v$-th term} in the Lagrangian (\ref{Lagrangian3rdorder}) are obtained from (\ref{threecurrentdefn}) where,
\small
\begin{eqnarray}\label{threecurr}
j_{ijk}&=&i\frac{\partial\mathcal{L}}{\partial(\partial_{i}\partial_{j}\partial_{k}\Phi)} \Phi +c.c. \nonumber \\
&=&i\bigg(|\Phi|^{4}\Phi\partial_{i}\partial_{j}\partial_{k}\Phi^*+2\Phi^3\partial_{i}\Phi^*\partial_{j}\Phi^*\partial_{k}\Phi^*-|\Phi|^{2}\Phi^2(\partial_{i}\Phi^*\partial_{j}\partial_{k}\Phi^*+\partial_{j}\Phi^*\partial_{k}\partial_{i}\Phi^*+\partial_{k}\Phi^*\partial_{i}\partial_{j}\Phi^*) \bigg) +c.c \nonumber \\ 
\end{eqnarray}
\normalsize
and likewise for other pieces in the Lagrangian.

\section{Generalisation to n-pole symmetries}
We now seek for a generalisation of our analysis which was carried out in the preceding sections to include ($n-1$)th order spatial shift symmetries,
\begin{equation}\label{shift-nth}
    \alpha=\alpha_0+\alpha_ix_i+\frac{1}{2}\alpha_{ij}x_ix_j+......+\frac{1}{n-1}\alpha_{i_1...i_{n-1}}x_{i_1}...x_{i_{n-1}}
\end{equation}

Then the general structure of  Noether currents for a theory with $\mathcal{L}(\Phi, \partial_{0}\Phi, \partial_{l_1}\Phi, \partial_{l_1}\partial_{l_2}\Phi, \partial_{l_1}\partial_{l_2}...\partial_{l_n}\Phi)$
is  given in the following form,
\begin{eqnarray}\label{conditionN}
\alpha j_{l_1}&-& (\partial_{{l_2}}\alpha)j_{l_1l_2} 
+(\partial_{l_2}\partial_{l_3}\alpha)j_{l_1l_2l_3}
+...+(-1)^{n-1}(\partial_{l_2}\partial_{l_3}..\partial_{l_n}\alpha)j_{{l_1}{l_2}..{l_n}}\nonumber\\ &=&\frac{\partial\mathcal{L}}{\partial(\partial_{l_1}\Phi)}\delta \Phi+2\frac{\partial\mathcal{L}}{\partial(\partial_{l_1}\partial_{l_2}\Phi)}\partial_{l_2}\delta \Phi+..+n\frac{{\partial\mathcal{L}}}{\partial(\partial_{l_1}\partial_{l_2}..\partial_{l_n}\Phi)}\partial_{l_2}\partial_{l_3}..\partial_{l_{n}}\delta \Phi \nonumber \\
&-&\partial_{l_2}\bigg(\frac{\partial\mathcal{L}}{\partial(\partial_{l_1}\partial_{l_2}\Phi)}\delta \Phi +.....+(n-1)\frac{{\partial\mathcal{L}}}{\partial(\partial_{l_1}\partial_{l_2}..\partial_{l_n}\Phi)}\partial_{l_3}..\partial_{l_n}\delta \Phi\bigg) \nonumber \\
&+&\partial_{l_2}\partial_{l_3}\bigg(\frac{\partial\mathcal{L}}{\partial(\partial_{l_1}\partial_{l_2}\partial_{l_3}\Phi)}\delta\Phi+.....+(n-2)\frac{{\partial\mathcal{L}}}{\partial(\partial_{l_1}\partial_{l_2}..\partial_{l_n}\Phi)}\partial_{l_4}..\partial_{l_n}\delta \Phi\bigg) \nonumber \\
&+&.....+(-1)^{n-1}\partial_{l_2}\partial_{l_3}..\partial_{l_n}\bigg(\frac{\partial\mathcal{L}}{\partial(\partial_{l_1}\partial_{l_2}..\partial_{l_n}\Phi)}\delta\Phi\bigg)+c.c. \nonumber \\
\end{eqnarray}
\normalsize
which is a natural extension of the $n=2$ (\ref{two-currdefn}) and $n=3$ (\ref{Noether3rd}) cases.
All tensor currents of rank less than $n$ are  derivable from the highest rank $n$ tensor current,
\begin{equation}\label{n-tensordefn}
j_{l_1}=\partial_{l_2}\partial_{l_3}..\partial_{l_{n}} j_{l_1l_2...l_{n}}\,\,;\,\,j_{l_1l_2}=\partial_{l_3}..\partial_{l_n}j_{{l_1}{l_2}..{l_n}}\,\,;\,\,j_{l_1l_2l_3}=\partial_{l_4}..\partial_{l_n}j_{{l_1}{l_2}..{l_n}} \,\,;\,\, .......
\end{equation}
where the $n$-th rank fully symmetric tensor current is given by,
\begin{equation}
j_{l_1l_2...l_{n}}=(-1)^{n-1}i\frac{{\partial\mathcal{L}}}{\partial(\partial_{l_1}\partial_{l_2}...
\partial_{l_{n}}\Phi)} \Phi
\end{equation}
The multiple conservation laws now follow from the following ``master equation",
\begin{eqnarray}\label{Mastereqn}
\partial_0(\alpha j_0)-\partial_i\bigg(\alpha j_i-(\partial_{{l_2}}\alpha) j_{il_2} 
+(\partial_{l_2}\partial_{l_3}\alpha) j_{il_2l_3} +...+(-1)^{n-1}(\partial_{l_2}\partial_{l_3}..\partial_{l_n}\alpha)j_{{i}{l_2}..{l_n}}\bigg)=0 \nonumber \\
\end{eqnarray}
If we factor out $\alpha$ then there there is a pairwise cancellation of the various terms, effected by exploiting 
(\ref{n-tensordefn}) and
we are left with the charge conservation law,
\begin{eqnarray}\label{usualchargeconservation}
\alpha(\partial_0j_0-\partial_ij_i)=0
\end{eqnarray}
Now, note that this procedure hides the additional conservation laws which can be  uncovered, as before, by opening up $\alpha$  using (\ref{shift-nth}) and inserting  in the ``master equation" (\ref{Mastereqn}) to yield (\ref{usualchargeconservation}) together with the following tower of ($n-1$) additional conservation laws,
\begin{eqnarray}\label{dipoleconservationn}
\partial_0J_0^k-\partial_iJ_i^k=0
\end{eqnarray}
\begin{eqnarray}
\partial_0J_0^{jk}-\partial_iJ_i^{jk}=0
\end{eqnarray}
\begin{eqnarray}
.....................=0
\end{eqnarray}
\begin{eqnarray}\label{n-poleconservation}
\partial_0J_0^{l_1l_2..l_{n-1}}-\partial_iJ_i^{l_1l_2..l_{n-1}}=0
\end{eqnarray}
obtained by separately equating the coefficients of $\alpha_0$, $\alpha_k$, $\alpha_{jk}$ till $\alpha_{l_1l_2..l_{n-1}}$ respectively. 
As an illustration the dipole conservation law for a system exhibiting the n-pole symmetry in the present formulation is given by (\ref{dipoleconservationn}) where the currents are defined in (\ref{newcurrents}). The rank 2 current $j_{ik}$ appearing there is obtained from (\ref{n-tensordefn}). 

%

\section{Conclusions}

Theories with polynomial space dependent shift symmetries, originally introduced in \cite{horava}, have recently been studied extensively due to their wide applications in various branches of physics, especially in unravelling and understanding features brought about by the imposition of new conservation laws that go beyond  usual charge conservation \cite{RB1, wang, yuan, hoyos, kristan}. These conservation laws were usually obtained by certain assumptions or by inspection. 

Polynomial space dependent shift symmetries involve, apart from  standard invariance under global (constant) transformations yielding charge conservation,  higher $n$-pole conservation laws. A natural question is whether  Noether's first theorem, that is strictly valid for global symmetries, can be extended to include such space dependent shift symmetries.

In this paper we have provided an appropriate generalisation of Noether's first theorem. A definition has been given  that yields all the $n$-pole currents. Also, a master equation has been provided from which various conservation laws associated with these currents are derived. The case of quasi-invariance where the Lagrangian changes by a boundary term under the transformation has also been accommodated.

A consequence of these shifted symmetries is the occurrence of higher derivative theories. This has raised issues in the literature \cite{hoyos} regarding their incorporation within the conventional fold of field theories. The successful generalisation of Noether's theorem presented in this paper would be a welcome step in this direction.

As a future possibility we hope to extend the present formulation to Noether's second theorem by carrying out a suitable gauging of the theories given here.

\section{Acknowledgements}

This work has been supported by a (DAE) Raja Ramanna Fellowship.

\end{document}